\documentclass[pra,twocolumn,showpacs]{revtex4}

\usepackage[latin1]{inputenc}
\usepackage[T1]{fontenc}
\usepackage{amsmath,amssymb,amsfonts,amsthm}
\usepackage{epsfig}
\usepackage{color}
\usepackage{xspace}
\usepackage{hyperref}
\usepackage{longtable}
\usepackage{my-style}

\newtheorem{theorem}{Theorem}
\newtheorem{corollary}{Corollary}

\begin{document}
\title{Two-party Models and the No-go Theorems}
\date{\today}
\author{Minh-Dung Dang}
\affiliation{GET-ENST \& LTCI-UMR 5141 CNRS, 46 rue
Barrault, 75634 Paris Cedex 13, France}

\begin{abstract}
In this paper, we reconsider the communication model used in the no-go
theorems on the impossibility of quantum bit commitment and oblivious
transfer. We state that a macroscopic classical channel may not be
replaced with a quantum channel which is used in the reduced model
proving the no-go theorems. We show that in some restricted cases, the
reduced model is insecure while the original model with a classical
channel is secure.
\end{abstract}


\maketitle


\section{\label{sec:introduction}Introduction}
The theorems on the \emph{security} of quantum key
distribution~\cite{LC99,SP00}, following the first protocol of Bennett
and Brassard~\cite{BB84}, and the theorems on the \emph{insecurity} of
two-party quantum bit commitment~\cite{LC97,May97,Bub01}, oblivious
transfer and general two-party quantum secure computation~\cite{Lo97}
are among the most interesting subjects in the field of quantum
cryptography.

In a bit commitment protocol, Alice sends the commitment information
about a secret bit to Bob, who cannot discover the bit, and when Alice
is supposed to reveal the bit, Bob can detect if Alice changes the
value of the bit.  Oblivious transfer is a computation protocol where
Alice enters two secret bits, Bob enters a choice to gain only one of
these while Alice cannot know Bob's choice. Coin flipping is a
protocol for Alice and Bob sharing a fairly random bits, i.e. none of
the parties can affect the probability distribution of the outcome.

These three primitives are used to construct secure computation for
generic two-party functions~\cite{Gol04book}. Classically, bit
commitment implements coin flipping and is implemented by oblivious
transfer~\cite{Kil88}, while oblivious transfer can be built from bit
commitment by transmitting quantum information~\cite{Cre94,Yao95}.

With the introduction of quantum information~\cite{Wie83},
cryptographers were willing to build unconditionally secure bit
commitment~\cite{BB84,BCJ+93}. But the first no-go theorem eliminated
these proposals~\cite{LC97,May97}. This knock alerted people about the
provable security of other two-party protocols, including key
distribution. While quantum key distribution is proved to be
secure~\cite{LC99,SP00}, two other no-go theorems were issued: quantum
secure two-party computation and so oblivious transfer are
impossible~\cite{Lo97}; quantum coin flipping with arbitrarily small
bias is impossible~\cite{Kit02}. We can add to the list another
no-go result: quantum bit commitment cannot be built from coin
flipping~\cite{Ken99}.

These no-go theorems execute the protocols on a quantum two-party
model. This model is proposed first by Yao, and consists of
a quantum machine on Alice side interacting with a quantum machine on
Bob side via a quantum machine~\cite{Yao95,Kre95}. The global states,
or \emph{images}, of a protocol are then described in a joint space of
Alice side $\hilbert{A}$, Bob side $\hilbert{B}$ and the quantum
channel $\hilbert{C}$. The states of the execution can be mixed
states, but all local random choice and computation made by Alice and
Bob can be purified and kept
at the quantum level~\cite{May97,LC97,Lo97}. The global state is then
described by a pure state in a larger space $\hilbert{A'}\otimes
\hilbert{C}\otimes \hilbert{B'}$ where $\hilbert{A'},\hilbert{B'}$ are
extended from $\hilbert{A},\hilbert{B}$ to purify the random choices
and measurements. Adopting the option that parts of the channel are
controlled by one of the two participants, the no-go theorems converts
the state of the joint computation (at a moment) into a bipartite
state of  joint space $\hilbert{A''}\otimes\hilbert{B''} =
(\hilbert{A'}\otimes\hilbert{CA})\otimes(
\hilbert{CB}\otimes\hilbert{B'})$ where $\hilbert{CA},\hilbert{CB}$
are the channel parts respectively controlled by Alice and Bob. In
such a bipartite space, bit commitment and oblivious transfer are
impossible.

In this paper, I try to criticize the application of the no-go
theorems to all possible two-party protocols, including
quantum-classical mixed protocols. In fact, a specified protocol with
classical messages would rather be implemented on a real physical
model with a classical channel. We state that a classical channel
can be macroscopic and cannot be controlled by any participants in
terms of purification. It should be viewed as a measurement which is
trusted by both Alice and Bob. I give some restricted cases where the
no-go theorems are valid with quantum communications but not with
quantum-classical mixed communications.

This critique is not to say that quantum bit commitment is
possible. Indeed, another reconsideration of bit commitment on a
complete model was made recently~\cite{AKS+06}, and we should wait for
reviews to confirm the results.

First, in Section \ref{sec:revision}, we do a revision on the no-go
theorems of Mayers, Lo \& Chau and the quantum model used in their
proofs. With the same arguments, in Section~\ref{sec:ext-nogo}, we
extend the no-go theorems with presence of a particular quantum
trusted third-party.  By this extension of the no-go theorems for a
particular quantum trusted third-party model, I show a result similar
to~\cite{Ken99}: bit commitment and oblivious transfer cannot be built
upon coin flipping with purely quantum communications.
Then, in Section~\ref{sec:reduction}, we
re-question about the arguments of Mayers, Lo \& Chau in reducing
general protocols from the mixed model with a classical channel to the
purely quantum model. In that section, we propose a generalized
model to adopt a real classical channel, and we give a particular
case-study where the no-go theorems are valid with purely quantum
communications but not with mixed communications.

In the following, we will use shortly ``protocols'' for all protocols
with quantum-classical mixed communications while ``quantum
protocols'' for  protocols with purely quantum communications.

\section{\label{sec:revision}Quantum model and no-go theorems}
\subsection{\label{sec:quantum-model}Quantum model}
In~\cite{Yao95}, for proving the security of bit-commitment-based
quantum oblivious transfer, Yao defined a quantum two-party protocol
as a pair of quantum machines interacting through a quantum
channel.

The protocol is then executed on a joint system consisting of Alice's
machine $\hilbert{A}$, Bob's machine $\hilbert{B}$, and the quantum
channel $\hilbert{C}$. Initially, each participant prepares a state
for its private system and the channel is in state $\ket{0}$.
The execution is alternating rounds of one-way
communications. For each round of one
participant $D \in \{A,B\}$, this performs a computation on the joint
space of his private system $\hilbert{D}$ and the messages
$\hilbert{C}$. The messages will be taken to the location of the
other for the next round. 

\begin{figure}[htb]
\resizebox{0.4\textwidth}{!}{\epsfig{file=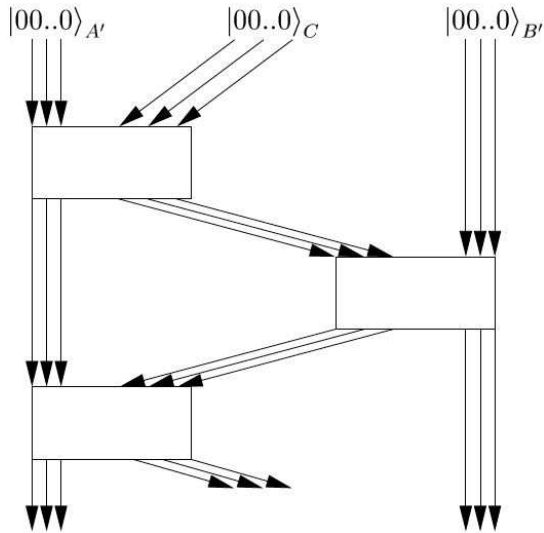}}
\caption{\label{fig:quantum-model}Quantum two-party model}
\end{figure}

This model is general for all quantum protocols, and has been widely
used for analyzing quantum protocols, e.g. the complexity of quantum
communication~\cite{Kre95,Wol02} and quantum interactive
proofs~\cite{Wat99}.

\subsection{\label{sec:nogo-theorems}No-go theorems}
The no-go theorems used the quantum model of Yao to prove the
insecurity of two-party protocols.

For protocols with  quantum-classical mixed communications, Yao said
``Although the above description is general enough to incorporate
classical computations and transmissions of classical information, it
is useful to separate out the classical parts in describing
protocols''~\cite{Yao95}. But his model was then widely used for all
quantum-classical mixed protocols. The arguments is that a classical
bit can be transmitted as a qubit. Mayers explicitly explained this
kind of \emph{quantum} communication for classical
information~\cite{May97}.

Because all random choices are made by either Alice or Bob, these
two participant can keep the random choices at the quantum level by
holding the entanglement with the quantum coins. Indeed, the fact that
a participant purifies or not his random variables does not change the
correctness of the protocol and the cheating strategy of
the other. All of local computations can be also kept at the quantum
level by delaying measurements to the final step. Thus, the global
\emph{images} of the the computation
can be described by a pure state lying on a larger space purifying all
local random variables and measurements:
$\hilbert{A'}\otimes\hilbert{C}\otimes\hilbert{B'}$. The execution of
the protocol is then a well specified unitary transformation
of the input state into the output state, and surprisingly
\emph{deterministic}.

In the quantum model, at each moment between two rounds
of communication, any quantum part of the channel must lie on either
Alice or Bob side. Thus, the global image at any moment can be
considered as a state lying in a bipartite space
$\hilbert{A''}\otimes\hilbert{B''}  =
\hilbert{A'}\otimes\hilbert{CA}\otimes\hilbert{CB}\otimes\hilbert{B'}$
where $\hilbert{CD}$ is for the channel part held by participant $D
\in \{A,B\}$.

Then, at the considered moment, the partial images of the computations
on Alice and Bob sides are
$$
\rho^{A''} = tr_{B''}(\rho), \quad \rho^{B''} = tr_{A''}(\rho),
$$
where $\rho$ is the global image which is a pure state in the global
space $\hilbert{A''}\otimes\hilbert{B''}$.

For the security of bit commitment on Bob side, the partial images on
Bob side at the moment before the opening phase must identical for the
commitments of $0$ and $1$:
$$
tr_{A''}(\rho_0) = tr_{A''}(\rho_1)\cdot
$$
The no-go theorem on bit commitment is issued by a theorem that states
that, in such a case, there exists a local unitary transformation on
$\hilbert{A''}$ that transforms $\rho_0$ to $\rho_1$. Alice can then
switch the computation before opening the secret bit~\cite{May97,LC97}.

For the security of one-sided computation, Lo discovered that if the
protocol is secure against Alice, then Bob has a local
unitary transformation, independent from Alice's input, that helps Bob
to learn the results computed from all of possible local
inputs~\cite{Lo97}.

We revise here Lo's theorem for one-sided computation. In fact, to
compute $f(i,j)$, Alice and Bob run together a unitary $U$
transformation on Alice's input $\ket{i}:i \in \{i_1,..,i_m\}$ joint
with Bob's input $\ket{j}:j \in \{j_1,..,j_n\}$. Other known local
variables can be omitted without generalization. At the end, Bob can
learn the result from the output state $\ket{v_{ij}} =
U(\ket{i}_A\otimes\ket{j}_B)$. But Alice can entangle her input $A$
with a private quantum dice $P$, i.e. prepares the initial state
$\frac{1}{\sqrt{n}}\sum_i \ket{i}_P\otimes\ket{i_A}$.

If Bob inputs $j_1$ then the initial state for the protocol is
\begin{equation}\label{equa:lo-initial}
\ket{u'}_{in} = \frac{1}{\sqrt{n}}\sum_i
\ket{i}_P\otimes\ket{i_A}\otimes\ket{j_1}_B,
\end{equation}
and at the end, the output state is
$$
\ket{v_{j_1}} = \frac{1}{\sqrt{n}}\sum_i
\ket{i}_P\otimes U (\ket{i_A}\otimes\ket{j_1}_B)\cdot
$$
Similarly, if Bob inputs $j_2$ then the output state is
$$
\ket{v_{j_2}} = \frac{1}{\sqrt{n}}\sum_i
\ket{i}_P\otimes U (\ket{i_A}\otimes\ket{j_2}_B) \cdot
$$
For the security on Alice side, the partial images must be identical,
i.e.
$$
tr_B(\projection{v_{j_1}}{v_{j_1}}) =
tr_B(\projection{v_{j_2}}{v_{j_2}})
$$
and then, there exists a local unitary transformation $U^{j_1,j_2}$ on
Bob local system such that
$$
\ket{v_{j_2}} = U^{j_1,j_2}\ket{v_{j_1}}\cdot
$$
Therefore, because $_P\bracket{i}{v_j} =
\frac{1}{\sqrt{n}}\ket{v_{ij}}$, the transformation $U^{j_1,j_2}$ is
universal for all Alice input $i$:
$$
\ket{v_{ij_2}} = U^{j_1,j_2}\ket{v_{ij_1}}\cdot
$$
Bob can enter $\ket{j_1}$, computes $\ket{v_{ij_1}}$ and measures it
to learn $f(i,j_1)$. However, to enable Bob to \emph{unambiguously}
get the result, $\ket{v_{ij_1}}$ must not be perturbed by his
measurement. Bob can transform it to $\ket{v_{ij_2}}$ by
$U^{j_1,j_2}$, measures to learn $f(i,j_2)$, and so on.

More strongly, imperfect protocols are also banned from reaching an
arbitrarily high security. There exist always a trade-off between the
security on one side and the insecurity on the other
side~\cite{SR01b}.

\section{\label{sec:ext-nogo}Extensions of no-go theorems for the
quantum model} 
\begin{figure}[htb]
\resizebox{0.4\textwidth}{!}{\epsfig{file=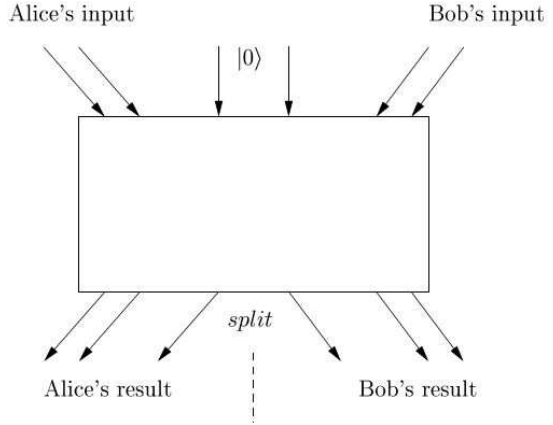}}
\caption{\label{fig:trusted-computation}The quantum trusted third-party}
\end{figure}

In this section, we suppose an honest third-party that
helps Alice and Bob to do some computations. We define a quantum
trusted third-party as a quantum device that can help Alice
and Bob to do any required computation. The third-party can used some
local pure variables for the computations. The local variables of the
trusted party are initialized to $\ket{0}$. At the end of the required
computation, the third-party splits all of the outputs, included the
local variables, into two parts, redirects one part to Alice, and one
part to Bob, cf. figure~\ref{fig:trusted-computation}. The execution
time of the computation done by the third-party is a elementary unit,
and we can consider as it immediately returns the results to the
participants.

With such a trusted third-party, we can extend the results of
\cite{May97,LC97,Lo97} for the quantum model:

\begin{theorem}[Extension of no-go theorem on bit commitment]
\label{thm:ext-bc-nogo}
Quantum bit commitment is insecure even with help of the specified quantum
trusted third-party.
\end{theorem}

\begin{theorem}[Extension of no-go theorem on secure computations]
\label{thm:ext-ot-nogo}
Quantum two-party secure computations are insecure even with help of
the specified quantum trusted third-party.
\end{theorem}

In fact, when the third-party uses only pure states as local input,
and immediately, splits and sends all of the qubits which participate
to the computations to Alice and Bob, the global state at any considered
moment is in some known pure state, according to the algorithm, in a
bipartite space relating only Alice and Bob sides. Therefore, the
no-go theorems remain valid.

For example, to prove the later theorem for one-sided secure
computation. We start with equation
\eqref{equa:lo-initial}. Attaching a pure state $\ket{0}_{A'B'}$,
locally prepared by the third-party, the initial state is
$$
\ket{u'}_{in} = \frac{1}{\sqrt{n}}\sum_i \ket{i}_P \otimes \ket{i}_A
\otimes \ket{j_1}_B \otimes \ket{0}_{A'B'}\cdot
$$
At the end of the computation, with help of the third-party, the
combined system is in state
$$
\ket{v_{j1}} = \frac{1}{\sqrt{n}}\sum_i \ket{i}_P \otimes U(\ket{i}_A
\otimes \ket{j_1}_B \otimes \ket{0}_{A'B'})
$$
where system $A'$ is set to $A$, system $B'$ is set to $B$ after the
split. Therefore, the remaining arguments of Lo's proofs can be
followed, cf. Section \ref{sec:nogo-theorems}.

\section{\label{sec:cf-based}Is Coin Flipping weaker than Bit Commitment?}
As a corollary of the extensions of the no-go theorems, cf. Section
\ref{sec:ext-nogo}, we conclude that
\begin{corollary}
In the quantum model, Coin Flipping is weaker than Bit Commitment and
Oblivious Transfer.
\end{corollary}
In \cite{Ken99}, Kent shown a similar result. In his paper, he
established a relativist model to implement coin flipping. With an
assumed quantum trusted party, we made the model more
comprehensible from a non-relativist point of view.

It's because we can suppose a trusted third-party that creates a
pair of qubits in Bell state $\ket{\Phi+}$ and sends each part to an
user. In such a model, coin flipping is realizable while bit commitment
and oblivious transfer are not with quantum communication, as shown by
Theorems \ref{thm:ext-bc-nogo}, \ref{thm:ext-ot-nogo}.

\section{\label{sec:reduction}Quantum model is not to prove the
insecurity} \subsection{\label{sec:mixed-model}Quantum-classical mixed
model} What is the difference between the two channel? A quantum
channel is a medium that we can used to transmit directly a quantum
state without disturbing it while a classical channel permits only one of
two discrete values for a classical bit. For example a macroscopic
electrical wire with tension $+5V$ for $0$ and $-5V$ for $1$.  It is
natural that in reality, a classical channel is well coupled with the
environment, and the decoherence is so strong that only $\ket{0}$ or
$\ket{1}$ is accepted, in terms of quantum information.

Imagine that in the specification of a protocol, at a certain
moment, a party has to measure some results of its quantum computation
and send the resulting classical messages to the other party. With a
macroscopic classical channel, the measurement must be carried
out. The sender can also memorize the emitted message. It is
convenient to see the classical channel as a trusted measurement
machine: the sender sends the qubits to the machine that measures,
doubles the output state, which is an eigenstate and cloneable, sends
one copy to the receiver and one copy back to the sender for
memorizing it.

Yao's model should be generalized for two-party protocols as a pair of
quantum machines interacting through a quantum channel \emph{and
necessarily a classical channel}. The model is also alternating rounds
of one-way communications. In each round, a participant $D \in
\{A,B\}$ performs a computation on the joint space of his private
system $\hilbert{D}$, the quantum messages $\hilbert{C}$ the classical
messages $\hilbert{R_M,D}$ received from the trusted measurement
machine $M$, and the messages $\hilbert{S_M,D}$ to be sent to the
measurement machine for producing classical messages.

For simplifying, the measurement machine should not copy the
output. This task can be carried out by the sender. In fact, measuring
a state $a\ket{0} + b\ket{1}$, the above machine produces $\ket{00}$
or $\ket{11}$. Instead, the sender can create $a\ket{00} + b\ket{11}$,
sends the first qubit to the machine that measures it and sends the
output state to the receiver. By this way, the sender keep a copy of
the measurement.

\begin{figure}[htb]
\resizebox{0.47\textwidth}{!}{\epsfig{file=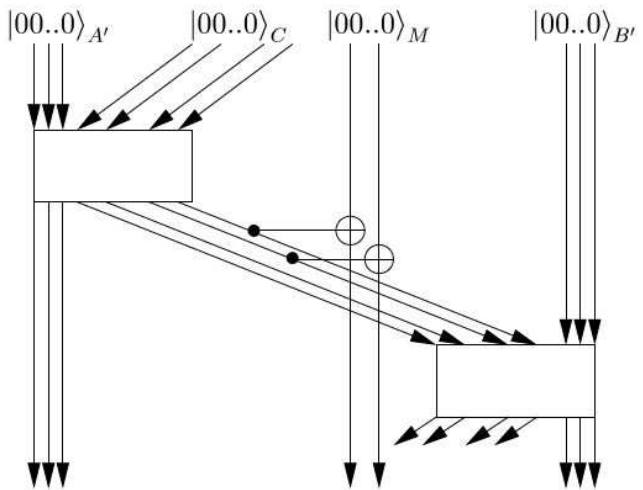}}
\caption{\label{fig:mixed-model}Mixed model}
\end{figure}

Therefore, the model is simplified, and consists of two machines
$\hilbert{A},\hilbert{B}$, a quantum channel $\hilbert{C}$ for both
quantum and classical messages and a trusted measurement machine $M$
with ancillas $\hilbert{M}$. The measurement is in fact a CNOT-like
gate whose controlling inputs are in the space of the sender's
``classical messages'' and targets are ancillas in the macroscopic
environment space $\hilbert{M}$, cf. figure~\ref{fig:mixed-model}.

In each communication round, a participant $D \in \{A,B\}$ does an
unitary computation on $\hilbert{D'}\otimes\hilbert{C}$ where
$\hilbert{D'}$ is extended from $\hilbert{D}$ to purify local
variables and measurements; the trusted machine applies the CNOT gate
to the ``classical messages'' in $\hilbert{C}$ and the environment of
the classical channel $\hilbert{M}$. The quantum messages and
``classical messages'' in $\hilbert{C}$ of the round are taken to the
other location for the next round. $\hilbert{M}$ is not controlled by
any participant.

\subsection{\label{cogent-reduction}Is the reduction cogent?}
It is obvious that all protocols can be \emph{correctly} implemented
on the quantum model by replacing classical bits by qubits. By this
way, a protocol is secure if its simulating quantum protocol is
secure. It's reasonable to use simulating quantum protocols \emph{to
prove the security of original protocols}~\cite{Yao95}. But,
vice-versa, it is not evident. Unfortunately, the no-go theorems used
the simulating quantum protocols \emph{to prove the insecurity of
original protocols}.

\begin{figure*}[htb]
\resizebox{0.85\textwidth}{!}{\epsfig{file=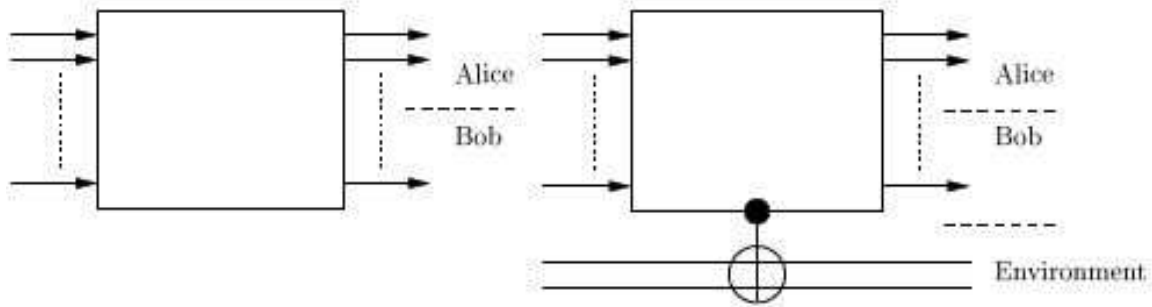}}
\caption{\label{fig:quantum-vs-mixed}Quantum vs. Mixed models}
\end{figure*}

For example, in the specification of a mixed protocol, a
participant makes the measurement of its computation to produce a
classical message, sends the messages via a the classical channel
which reproduces a corresponding state in the computation basis for
the other participant. In the model of the specification with a
macroscopic channel for sending classical message, the receiver really
receives one of the eigen-states of the measurement made by the sender.

While, by simulating with a quantum channel, the sender
equivalently creates a quantum mixed state as the sum of the above
measurement eigen-states weighted by the corresponding
probabilities. However, the sender can prepare this mixed state as
part of a bipartite pure entangled state. We see this by citing to an
early simple example: at a certain moment, a participant gets
$a\ket{0} + b\ket{1}$, measures it and sends the result, either
$\ket{0}$ with probability $a^2$ or $\ket{1}$ with probability
$b^2$. This classical message is described in terms of quantum
information as $a^2\projection{0}{0}+b^2\projection{0}{0}$. With a
quantum channel, the sender can send $(a\ket{0} +
b\ket{1})\otimes\ket{0}_C$ through a CNOT gate to make
$(a\ket{0}\otimes\ket{0}_C  + b\ket{1}\otimes\ket{1}_C)$ and send
qubit $C$ to the receiver.

The above quantum communication of classical messages gave to the
participants an extra entanglement that does not exist in the
specification of the protocol with mixed communications. Indeed, this
entanglement could be used as a powerful attack.  We will explicitly
expose this with a case-study where, if the receiver used the received
message to do some quantum computation and sends back the result, the
sender could learn more information with entanglement attack by the
effect of \emph{super-dense coding}~\cite{BW92}.

Such an bipartite entanglement should be destroyed by the measurement
of the classical channel. In our model for the classical channel, the
sender has to send qubit $C$ to the measurement machine of the
classical channel that makes instead the state
$(a\ket{0}\otimes\ket{0}_M\otimes\ket{0}_C  +
b\ket{1}\otimes\ket{1}_M\otimes\ket{1}_C)$ and sends qubit $C$ to the
receiver.

From a global view, one can see that if the two parties follow a
protocol using a defined classical channel, the
global system should lie in a tripartite spaces joining
$\hilbert{A}\otimes\hilbert{C,A}$, $\hilbert{B}\otimes\hilbert{C,B}$
and $\hilbert{M}$. The crucial argument that the whole system is
described in a bipartite space,  used in the proofs of the no-go
theorems~\cite{May97,LC97,Lo97}, is not valid in the mixed
model, cf. figure~\ref{fig:quantum-vs-mixed}.

All of the proposed quantum bit commitment protocols fall into the
quantum model~\cite{BB84,BCJ+93,Yue00}, where all communications are
realized with quantum messages except the final step to open the
committed bit. Then, they are directly attacked by the no-go
theorems~\cite{May97,LC97,Bub01}.

Maybe, unconditionally secure bit commitment is also impossible in the
mixed model, as concluded by a recent study in a preprint paper of
Mauro d'Ariano et al.~\cite{AKS+06},  but the arguments used in Mayer
and Lo \& Chau proofs for all possible protocols are not evident.

We think that, the insecurity of a protocol should be considered when
implementing it on a quantum-classical mixed model that match better
the real world. Normally, a macroscopic classical channel is coupled
with a trusted environment and measures the quantum messages sent
through it. We expect that, to be more convincing, no-go theorems
should be proved for the mixed model we proposed in the previous
section.

\subsection{\label{sec:honest-oot-gate} Case-study: an honest
third-party O-OT gate}

Let verify a quantum oblivious transfer protocol with a pure trusted
party, cf. \ref{sec:ext-nogo}. In our protocol, the pure trusted
party uses tree local qubits. For simplifying, we initialize the first
and the second qubit to be entangled and in state $\ket{\Phi+}$. In
fact, the trusted party can prepare this state from $\ket{00}$ by a
doing $\pi/2$ rotation on the first qubit and sending the two qubit
through CNOT gate whose target is the second qubit. The third qubit is
initialized to $\ket{0}$.

Inspired from Bennett et al.~\cite{BDS+96}, we use the notations:
\begin{align*}
\widetilde{00} = \ket{\Phi+} = (\ket{00} + \ket{11})/\sqrt{2},\\
\widetilde{01} = \ket{\Phi-} = (\ket{00} - \ket{11})/\sqrt{2},\\
\widetilde{10} = \ket{\Psi+} = (\ket{01} + \ket{10})/\sqrt{2},\\
\widetilde{11} = \ket{\Psi-} = (\ket{01} - \ket{10})/\sqrt{2}\cdot
\end{align*}

\begin{figure}[htb]
\resizebox{0.5\textwidth}{!}{\epsfig{file=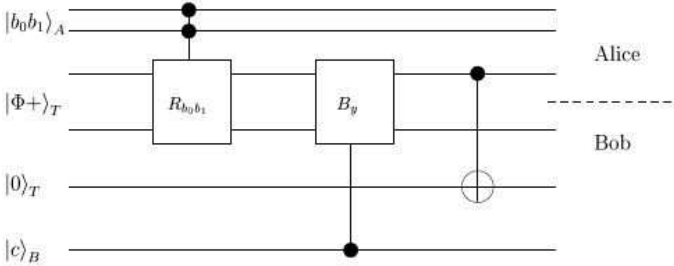}}
\caption{\label{fig:oot-gate}A third-party gate for O-OT protocol}
\end{figure}

Let $b_0,b_1$ be the two bits that Alice want to send and $c$ be Bob's
choice. The trusted party does a controlled $\pi$ rotation $R_{b_0b_1}$
on the first qubit, according to $b_0,b_1$:
$$
R_{00} = I, R_{01} = \sigma_z, R_{10} = \sigma_x, R_{11} = \sigma_y\cdot
$$
The first and second qubits are obtained in state
$\widetilde{b_0b_1}$. Next, the trusted party applies the bilateral
$\pi/2$ rotation $B_y$ to the first and second qubits in case
$c=1$~\cite{BDS+96}:
\begin{align*}
\widetilde{00} \rightarrow_{B_y} \widetilde{00},\\
\widetilde{01} \rightarrow_{B_y} \widetilde{10},\\
\widetilde{10} \rightarrow_{B_y} \widetilde{01},\\
\widetilde{11} \rightarrow_{B_y} \widetilde{11}\cdot
\end{align*}
The trusted party applies then the CNOT gate whose controlling input is
the first qubit and the target is the third qubit. Finally, the
trusted party splits the outputs, sends back Alice's qubits
with his first local qubit to Alice, and sends back Bob's
qubit with its second and third local qubits to Bob,
cf. figure~\ref{fig:oot-gate}.

In case Alice and Bob communicate with the trusted party via quantum
channels, they can send directly quantum inputs. The computation of
the trusted party is a quantum circuit acting on
$6$ qubits: two for Alice's inputs, tree for the local qubits, one for
Bob's input. Alice can prepare a superposition
$$
\frac{1}{2}(\ket{00} + \ket{01} + \ket{10} + \ket{11})\cdot
$$
The global input state is then
$$
\ket{in} = \frac{1}{2}(\ket{00}_A + \ket{01}_A + \ket{10}_A +
\ket{11}_A)\ket{\Phi+}_T\ket{0}_T\ket{c}_B\cdot
$$
If Bob sends $\ket{c} = \ket{0}$ then the computation is
\begin{align*}
\ket{in} & \rightarrow_{R_{b_0b_1}} & \frac{1}{2} [ &
\ket{00}_A\widetilde{00}_T + \ket{01}_A\widetilde{01}_T\\
	&&& + \ket{10}_A\widetilde{10}_T +
\ket{11}_A\widetilde{11}_T]\ket{0}_T\ket{0}_B\\
	&\rightarrow_{B_y} & \frac{1}{2\sqrt{2}}[ &
\ket{00}_A(\ket{00}_T+\ket{11}_T)\\
	&&& + \ket{01}_A(\ket{00}_T - \ket{11}_T)\\
	&&& + \ket{10}_A(\ket{01}_T+\ket{10}_T) +\\
	&&&\ket{11}_A(\ket{01}_T - \ket{10}_T)]\ket{0}_T\ket{0}_B\\
	&\rightarrow_{CNOT} &  \frac{1}{2\sqrt{2}}[ &
\ket{00}_A(\ket{000}_T+\ket{111}_T)\\
	&&&+ \ket{01}_A(\ket{000}_T - \ket{111}_T)\\
	&&& + \ket{10}_A(\ket{010}_T+\ket{101}_T)\\
	&&& + \ket{11}_A(\ket{010}_T - \ket{101}_T)]\ket{0}_B\\
	&\rightarrow_{split} &  \frac{1}{2\sqrt{2}}[ & (\ket{000}_A +
\ket{010}_A)\ket{000}_B\\
	&&& + (\ket{001}_A - \ket{011}_A)\ket{110}_B\\
	&&& + (\ket{100}_A + \ket{110}_A)\ket{100}_B\\
	&&& + (\ket{101}_A - \ket{111}_A)\ket{010}_B]\cdot
\end{align*}
If Bob sends $\ket{c}=\ket{1}$ then the computation is
\begin{align*}
\ket{in} & \rightarrow_{R_{b_0b_1}} & \frac{1}{2}[ &
\ket{00}_A\widetilde{00}_T + \ket{01}_A\widetilde{01}_T +\\
	&&& \ket{10}_A\widetilde{10}_T +
\ket{11}_A\widetilde{11}_T]\ket{0}_T\ket{1}_B\\
	&\rightarrow_{B_y} &  \frac{1}{2\sqrt{2}}[ &
\ket{00}_A(\ket{00}_T+\ket{11}_T)\\
	&&&+ \ket{01}_A(\ket{01}_T + \ket{10}_T)\\
	&&& + \ket{10}_A(\ket{00}_T-\ket{11}_T)\\
	&&& + \ket{11}_A(\ket{01}_T - \ket{10}_T)]\ket{0}_T\ket{1}_B\\
	&\rightarrow_{CNOT} & \frac{1}{2\sqrt{2}}[ &
\ket{00}_A(\ket{000}_T+\ket{111}_T)\\
	&&& + \ket{01}_A(\ket{010}_T + \ket{101}_T)\\
	&&& + \ket{10}_A(\ket{000}_T - \ket{111}_T)\\
	&&& +\ket{11}_A(\ket{010}_T - \ket{101}_T)]\ket{1}_B\\
	&\rightarrow_{split} &  \frac{1}{2\sqrt{2}}[ & (\ket{000}_A +
\ket{100}_A)\ket{001}_B\\
	&&& + (\ket{001}_A - \ket{101}_A)\ket{111}_B\\
	&&& + (\ket{010}_A + \ket{110}_A)\ket{100}_B\\
	&&& + (\ket{011}_A - \ket{111}_A)\ket{011}_B]\cdot
\end{align*}
We see that the reduced density matrices at Alice's location are
different for the two cases, and so $c$ is not secure against Alice.

\begin{figure}[htb]
\resizebox{0.5\textwidth}{!}{\epsfig{file=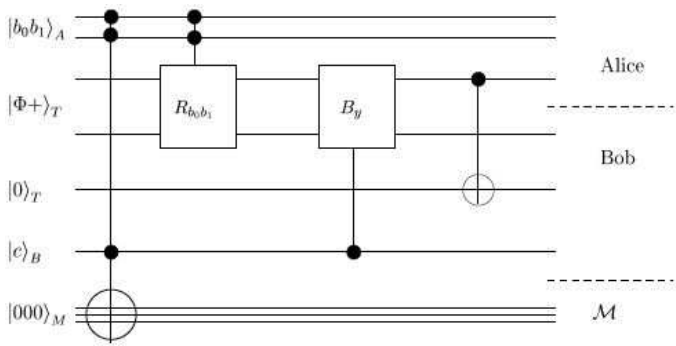}}
\caption{\label{fig:measure-oot-gate}Third-party gate for O-OT
protocol with classical channels}
\end{figure}

However, if Alice and Bob are subjected to send $b_0,b_1,c$ to $T$ via
classical channels, cf. figure \ref{fig:measure-oot-gate}. The inputs
will be measured and projected onto the computation basis. For example
on Alice side, simply speaking, Alice inputs $\ket{b_0b_1}$ can only
take one of the $4$ values $\ket{00}, \ket{01}, \ket{10}, \ket{11}$
and for any of these cases, Alice cannot discover $c$. Using the
defined model for the classical channel, Alice sends her inputs through
CNOT gates whose targets are in the measurement machine $M$ of the
classical channel. The output is entangled with $M$. The final states
of the computations for $c = 0$ and $c=1$ are
\begin{align*}
\ket{out_0} &=  \frac{1}{2\sqrt{2}}[ & (\ket{000}_A\ket{00}_M +
\ket{010}_A\ket{01}_M)\ket{000}_B\\
	&& + (\ket{001}_A\ket{00}_M - \ket{011}_A\ket{01}_M)\ket{110}_B\\
	&& + (\ket{100}_A\ket{10}_M + \ket{110}_A\ket{11}_M)\ket{100}_B\\
	&& + (\ket{101}_A\ket{10}_M - \ket{111}_A\ket{11}_M)\ket{010}_B],\\
\ket{out_1} &= \frac{1}{2\sqrt{2}} [ & (\ket{000}_A\ket{00}_M +
\ket{100}_A\ket{10}_M)\ket{001}_B\\
	&& + (\ket{001}_A\ket{00}_M - \ket{101}_A\ket{10}_M)\ket{111}_B\\
	&& + (\ket{010}_A\ket{01}_M + \ket{110}_A\ket{11}_M)\ket{100}_B\\
	&& + (\ket{011}_A\ket{01}_M - \ket{111}_A\ket{11}_M)\ket{011}_B]\cdot
\end{align*}
The reduced matrices of tree qubits at Alice location are gained by
tracing out $M$ part and $B$ part, and become $I/8$ for both two
values of $c$.

Similar analyses on Bob side shows that the protocol is also
secure. Therefore, with help of classical communications, the protocol
becomes secure on both sides.

\section{\label{conclusions}Summary}
Our arguments were based on the difference between communicating
classical information in a classical manner and in a quantum manner.

\begin{figure}[htb]
\resizebox{0.4\textwidth}{!}{\epsfig{file=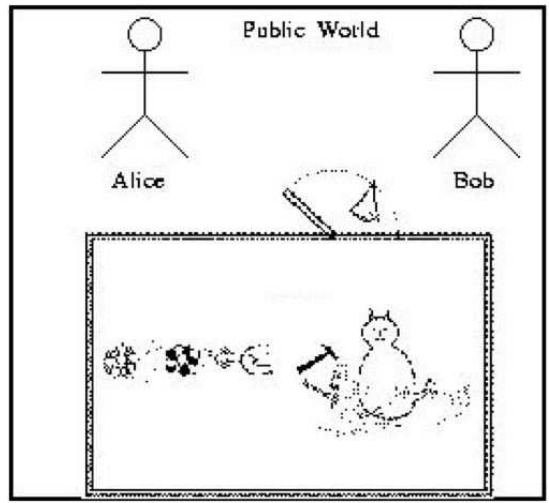}}
\caption{\label{fig:public-cat}Telling the cat's state via a
classical public environment}
\end{figure}

The discussions take us back to a similar problem of Schrodinger's
Cat~\cite{Gri04}. Imagine that Alice owns a Schrodinger Box, and at
a certain moment, has to tell Bob whether the cat is dead or alive.

\begin{figure}[htb]
\resizebox{0.4\textwidth}{!}{\epsfig{file=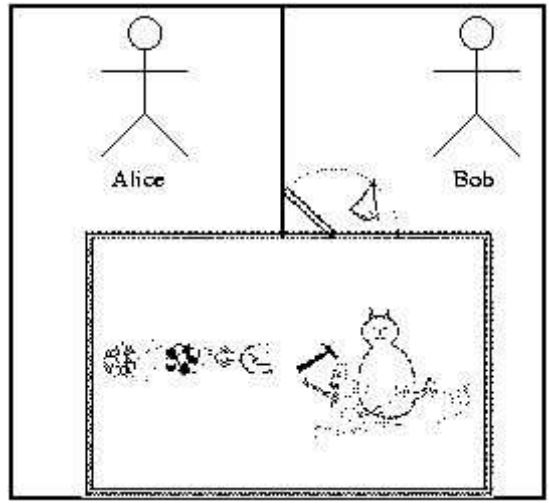}}
\caption{\label{fig:private-cat}Telling the cat's state via
private environments}
\end{figure}

If Alice and Bob live in a same public environment, e.g. in a same
room, Alice does this via a classical channel, e.g. the acoustic
channel: Alice has to open the box and sound what she sees. It is
equivalent to as though they open the box together,
cf. figure~\ref{fig:public-cat}. In another way, Alice can give the
box to Bob and let him open it. However, Bob can open the box in a
private environment. We can say that Alice and Bob live in two
separate quantum worlds. Imagine that Alice and Bob live in two
isolated rooms. Alice puts the observable hole of her box through the
wall into Bob's room, and the two rooms remain always
isolated, cf. figure~\ref{fig:private-cat}. It is as though Bob's
measurement devices are thrown to a private quantum space.

We see that, within the classical concepts, a classical message
transmitted from Alice to Bob must be ``comprehensible'' by both
parties in a same reference frame. It is due to a classical channel as
a common environment that both Alice and Bob refer to. Classical
information can be viewed as quantum information, but measured by this
reference environment. All quantum measurement devices making classical
messages in a protocol between Alice and Bob must be thrown to this
trusted Hilbert space. In this paper, we have used these concepts of
communication of classical information, as in the former case of
telling the cat's state.

Nevertheless, in the purely quantum model used by Mayers and Lo \&
Chau, there is no such a common space, and the measurement devices of
each party for making classical information are thrown to a private
Hilbert space of that party~\cite{May97}.

Another example is the difference between the result of the
tossing of a classical random bit and the splitting of the EPR pair
$\ket{\Phi+}$. We see that the result of tossing of a random bit is a
pair of $(0_A,0_B)$ or $(1_A,1_B)$ with equal probability $1/2$. In
terms of quantum information, the random bits are described by the
density matrix:
$$
r_{AB} = \frac{1}{2}\projection{0_A0_B}{0_A0_B} +
\frac{1}{2}\projection{1_A1_B}{1_A1_B}\cdot
$$
On the other hand, the EPR pair is
$$
\ket{\Phi+} = \frac{1}{\sqrt{2}}(\ket{0_A0_B} + \ket{1_A1_B})
$$
The two states are indeed different. The EPR pair really implements
the tossing only when Alice and Bob have a common reference, for
example measurement devices coupled with a common environment that
project each qubit to a same basis $\{\ket{0},\ket{1}\}$. This
measurement is trusted by both party, and used as a Hilbert space
$\hilbert{M}$ of reference, and the measurement devices can be thrown
to it. The measurement of the EPR pair gives
$$
\ket{\psi} = \frac{1}{\sqrt{2}}(\ket{0_A0_B0_M} + \ket{1_A1_B0_M}),
$$
and Alice and Bob get exactly a pair of random bits $r_{AB}$:
$$
tr_M(\projection{\psi}{\psi}) = \frac{1}{2}\projection{0_A0_B}{0_A0_B}
+ \frac{1}{2}\projection{1_A1_B}{1_A1_B}\cdot
$$

With the above concepts of communicating classical information, we
summarize that
\begin{itemize}
\item A general protocol, specified with classical and quantum
messages can be correctly implemented in the quantum model with only a
quantum channel. We can say that the original protocol is secure if
the simulating protocol in the quantum model is secure. However, we
have no right to use the simulating protocol in the quantum model to
prove the insecurity of the original protocol. We should consider its
insecurity in a model that would match better the real world with
quantum and classical channels.
\item We supposed that a classical channel is normally well coupled
with the environment and may not be controlled by neither Alice nor
Bob. It is convenient to see it as a trusted measurement which sends
back the classical outcomes to Alice and Bob.
\item We shown that in some special cases, the original protocol is
secure in presence of a classical channel while its simulating
protocol in the purely quantum model is insecure.
\end{itemize}

\bibliography{quantum-biblio,crypto-biblio}

\end{document}